\documentclass[aps,prd,superscriptaddress,twocolumn]{revtex4-1}

\usepackage{wallpaper}
\usepackage{bm}
\usepackage{amsmath}
\usepackage[colorlinks,linkcolor=blue,anchorcolor=blue,citecolor=red]{hyperref}
\usepackage{ulem}
\usepackage{multirow}

\begin{document}

\title{Asteroseismology of neutron stars with both hyperons and $\Delta$ resonances}
\author{Hao~Sun}
\affiliation{Center for Gravitation and Cosmology, College of Physical Science and Technology, Yangzhou University, Yangzhou 225009, China}

\author{Jia-Xing~Niu}
\affiliation{Center for Gravitation and Cosmology, College of Physical Science and Technology, Yangzhou University, Yangzhou 225009, China}

\author{Yong~Gao}
\affiliation{Max Planck Institute for Gravitational Physics (Albert Einstein Institute), 14476 Potsdam, Germany}

\author{Hong-Bo~Li}
\affiliation{Kavli Institute for Astronomy and Astrophysics, Peking University, Beijing 100871, China}

\author{Cheng-Jun Xia}
\email{cjxia@yzu.edu.cn}
\affiliation{Center for Gravitation and Cosmology, College of Physical Science and Technology, Yangzhou University, Yangzhou 225009, China}

\date{\today}

\begin{abstract}
Employing various relativistic energy density functionals for nucleon-nucleon interactions, we investigate the impact of hyperons and $\Delta$ resonances on the frequencies of non-radial oscillations in neutron stars, where the universal coupling scheme is adopted for the $\Delta$-meson couplings. It is found that the inclusion of $\Delta$ resonances is essential for neutron stars to accommodate the recent mass and radius measurements of PSR J0030+0451, PSR J0740+6620, PSR J0437-4715, PSR J0614-3329, and HESS J1731-347. As $\Delta$ resonances start to emerge in neutron stars' inner core, the $g$-mode oscillation energy becomes concentrated within the $\Delta$-admixed region, leading to a sharp increase in the $g$-mode frequency. We also examine the $f$-mode and $p$-mode oscillations and find that the impact of $\Delta$ resonances on these modes is less pronounced than the $g$-modes. The oscillation frequencies calculated in the Cowling approximation are then compared with those from full general relativity, confirming that the Cowling approximation introduces an error of approximately 20\% for the $f$-modes and within 10\% for the $g$-modes. The notable effect of $\Delta$ resonances on neutron stars' $g$-mode frequencies holds important implications for probing the internal composition of neutron stars.
\end{abstract}

\maketitle

\section{Introduction}
Due to the inability of solving QCD at large densities, the properties and composition of cold dense matter remains an open question~\cite{Dutra2012_PRC85-035201, Dutra2014_PRC90-055203, Baym2018_RPP81-056902, Xia2020_PRD102-023031, Xia2024_PRD110-114009}. Fortunately, being the products of gravitational collapse from massive stars, neutron stars provide unique astrophysical laboratories for probing the properties of ultra-dense matter under conditions far beyond those achievable in terrestrial experiments, where the observational masses and radii of pulsar-like objects could be employed to constrain the equation of states (EOS) of neutron star matter, e.g., those from the binary neutron star merger event {GRB} 170817A-{GW}170817-{AT} 2017gfo~\cite{LVC2018_PRL121-161101}, the pulse-profile modelings with the {NICER} and {XMM}-Newton data for PSR J0030+0451,  PSR J0740+6620, PSR J0437-4715, and PSR J0614-3329~\cite{Riley2019_ApJ887-L21, Riley2021_ApJ918-L27, Miller2019_ApJ887-L24, Miller2021_ApJ918-L28, Choudhury2024_ApJ971-L20, Mauviard2025_TAJ995-60}, as well as the central compact object (CCO) within the supernova remnant HESS J1731-347~\cite{Doroshenko2022_NA6-1444}.

Particularly, recent observations on PSR J0437-4715 ($M=1.42_{-0.04}^{+0.04}\ M_\odot$ and $R=11.36_{-0.62}^{+0.94}$ km)~\cite{Choudhury2024_ApJ971-L20} and PSR J0614-3329 ($M=1.44_{-0.07}^{+0.06}\ M_\odot$ and $R=10.29^{+1.01}_{-0.86}$ km)~\cite{Mauviard2025_TAJ995-60} suggest rather small radii for neutron stars, challenging the traditional neutron star model comprised of $npe\mu$ matter. At large densities, nevertheless, conventional nucleonic matter may turn into exotic states with non-nucleonic degrees of freedom, such as hyperons, $\Delta$ resonances, mesons, and quarks, which softens the EOS and consequently reduces the mass and radius of a neutron star~\cite{Togashi2016_PRC93-035808, Sun2019_PRD99-023004, Malfatti2020_PRD102-063008, Li2020_PLB810-135812, Marquez2022_PRC106-055801, Issifu2023_MNRAS522-3263, Gao2025_PRD112-083041, Grundler2025_PRD112-103012, Tang2025_PRD112-083009, Xia2025_IJMPA0-2550180, Albino2025, Kalita2025, Ayriyan2025}. Considering the possible existence of $\Delta$ resonances and hyperons in neutron stars, it was shown that the corresponding neutron star structures (masses, radii, and tidal deformabilities) could be consistent with various astrophysical observations~\cite{Sun2019_PRD99-023004, Malfatti2020_PRD102-063008, Li2020_PLB810-135812, Marquez2022_PRC106-055801, Issifu2023_MNRAS522-3263}.

Nevertheless, significant degeneracy remains among various neutron star models, where the emergence of other non-nucleonic degrees of freedom or varying the energy density functionals could also accommodate the observational neutron star structures~\cite{Togashi2016_PRC93-035808, Sun2019_PRD99-023004, Malfatti2020_PRD102-063008, Li2020_PLB810-135812, Marquez2022_PRC106-055801, Issifu2023_MNRAS522-3263, Gao2025_PRD112-083041, Grundler2025_PRD112-103012, Tang2025_PRD112-083009, Xia2025_IJMPA0-2550180, Albino2025, Kalita2025, Ayriyan2025}. It is thus favorable to adopt more discriminating probes to identify the matter contents of neutron stars. With increasing number and precision of gravitational wave detections~\cite{Luo2016_CQG33-035010, Hu2017_NSR4-685, Maggiore2020_JCAP2020-050}, the internal compositions of neutron stars may be unveiled with asteroseismology using the vast observational datasets. In particular, neutron stars exhibit characteristic eigenoscillations known as quasi-normal modes (QNMs)~\cite{Kokkotas1999_LRR2-1, Ho2020_PRD101-103009, Yu2017_MNTAS470-350}, where the non-radial oscillations can be classified into gravity modes ($g$-modes), pressure modes ($p$-modes), modes influenced by the Coriolis force ($r$-modes), pure spacetime modes ($w$-modes)~\cite{dey2025_arXiv, Kokkotas1992_MNRAS225-119}, and centrifugal modes in rotating stars. Both the fundamental $f$-mode and the composition-driven $g$-mode fall within the sensitivity range of current-generation gravitational wave detectors. Those modes may be excited resonantly by the strong tidal interactions during binary neutron star mergers~\cite{Lai1994_MNRAS270-611-629, Xu2017_PRD96-083005}, potentially imprinting detectable phase shifts on the gravitational waveform and may be observable by the upcoming detectors~\cite{Thorne1967_ApJ-149-591, Punturo2010_CQG27-194002, Regimbau2017_PRL118-151105, Abbott2017_CQG34-044001, Abbott2020_LRR23-3, Maggiore2020_JCAP2020-050}.

In fact, recent asteroseismology studies have begun to explore neutron stars containing $\Delta$ resonances~\cite{Parmar2025-PRD112-023016, Jyothilakshmi2025_sym17-230, Rather2023_PRD107-123022, Rather2025_PRD112-023013, Kalita2024_JCAP2024-65}, while the investigations on $g$-mode oscillations remain relatively scarce. We therefore further examine the $g$-mode oscillations in neutron stars including both hyperons and $\Delta$ resonances, which are particularly sensitive to the composition of dense matter. It was shown that with the emergence of quarks or hyperons, the $g$-mode oscillation frequencies of hybrid stars or hyperonic stars increase~\cite{Tran2023_PRC108-015803, Jaikumar2021_PRD103-123009, Wei2020_ApJ904-187, Constantinou2021_PRD104-123032, Zhao2022_PRD105-103025, Zheng2023_PRD107-103048}. This work extends such investigations to the cases including $\Delta$ resonances, where the impacts on $g$, $f$, and $p$-mode oscillations in neutron stars are examined. Note that our estimations on various oscillation modes are primarily performed adopting the Cowling approximation~\cite{Cowling1941_MNRAS101-367}, which introduces small deviations in the obtained frequencies. In general, for the $g$-modes of primary interest here, the Cowling approximation remains highly accurate. To illustrate the uncertainty introduced by employing the Cowling approximation, we also estimate the $g$, $f$, and $p$-mode frequencies of neutron stars in full general relativity, where the Cowling approximation introduces an error of approximately 20\% for the $f$-mode and within 10\% for the $g$-mode.

This paper is organized as follows. In Sec.~\ref{sec:the}, we present the relativistic-mean-field (RMF) models adopted here and the $\Delta$/hyperon-meson coupling constants. Sec.~\ref{sec:nonradial} outlines the methods for estimating nonradial oscillation frequencies, both in the Cowling approximation and in full general relativity, describes the numerical techniques employed, and presents a detailed discussion of our results, unveiling the mechanism for the increase in $g$-mode frequency due to $\Delta$ resonances. Finally, Sec.~\ref{sec:res} summarizes our findings and outlines prospects for future work. Throughout this paper, we use natural units $G=c=1$ and the metric signature $(-, +, +, +)$.

\section{\label{sec:the}Neutron star composition and structure}

\subsection{Neutron star matter}

To fix the EOSs of neutron star matter, we employ RMF model~\cite{Brockmann1977_PLB69-167, Boguta1981_PLB102-93, Mares1989_ZPA333-209, Mares1994_PRC49-2472, Toki1994_PTP92-803, Song2010_IJMPE19-2538, Tanimura2012_PRC85-014306, Wang2013_CTP60-479}, which has been successfully applied to describe finite (hyper)nuclei~\cite{Reinhard1989_RPP52-439, Ring1996_PPNP37_193-263, Meng2006_PPNP57-470, Paar2007_RPP70-R02, Meng2015_JPG42-093101, Typel1999_NPA656-331, Vretenar1998_PRC57-R1060, Lu2011_PRC84-014328} and baryonic matter~\cite{Ban2004_PRC69-045805, Weber2007_PPNP59-94, Long2012_PRC85-025806, Sun2012_PRC86-014305, Wang2014_PRC90-055801, Fedoseew2015_PRC91-034307, Gao2017_ApJ849-19}. Specifically, we adopt three relativistic energy density functionals (RDFs) for nuclear matter with density-dependent coupling constants, i.e., TW99~\cite{Typel1999_NPA656-331}, PKDD~\cite{Long2004_PRC69-034319} and DD-ME2~\cite{Lalazissis2005_PRC71-024312}. Beside nucleons, we consider also hyperons ($\Lambda^0$, $\Sigma^{+,0,-}$, and $\Xi^{0,-}$) and $\Delta$ baryons ($\Delta^{++}$, $\Delta^{+}$, $\Delta^{0}$, and $\Delta^{-}$) at high densities~\cite{Sun2019_PRD99-023004}. The interaction $\Delta$ between baryons is then mediated by three types of mesons ($\sigma$, $\omega$, and $\boldsymbol{\rho}$). The resulting Lagrangian density expression is then given by~\cite{Sun2019_PRD99-023004, Meng2016, Malik2017_PRC96-035803}
\begin{eqnarray}
\mathcal{L}
 &=& \sum_{\mathrm{b}} \bar{\psi}_\mathrm{b}
       \left[  i \gamma^\mu \partial_\mu- m_\mathrm{b} - g_{\sigma\mathrm{b}}\sigma
              - g_{\omega\mathrm{b}} \gamma^\mu \omega_\mu
         \right.
\nonumber \\
 &&\mbox{}\left.- g_{\rho\mathrm{b}} \gamma^\mu \boldsymbol{\tau}_\mathrm{b} \cdot \boldsymbol{\rho}_\mu
              - \gamma^\mu A_\mu q_\mathrm{b}
              \right] \psi_\mathrm{b} + \frac{1}{2}\partial_\mu \sigma \partial^\mu \sigma
\nonumber \\
 &&\mbox{}   - \frac{1}{2}m_\sigma^2 \sigma^2 - \frac{1}{4} \omega_{\mu\nu}\omega^{\mu\nu} + \frac{1}{2}m_\omega^2 \omega_\mu\omega^\mu
\nonumber \\
 &&\mbox{} - \frac{1}{4} \boldsymbol{\rho}_{\mu\nu}\cdot\boldsymbol{\rho}^{\mu\nu}
     + \frac{1}{2}m_\rho^2 \boldsymbol{\rho}_\mu\cdot\boldsymbol{\rho}^\mu - \frac{1}{4} A_{\mu\nu}A^{\mu\nu}
\nonumber \\
 &&\mbox{}
     +\sum_{l=e,\mu} \bar{\psi}_l \left[ i \gamma^\mu \partial_\mu - m_l + e \gamma^\mu A_\mu \right]\psi_l,
\label{eq:Lagrange_1}
\end{eqnarray}
where ${\tau}_n = -{\tau}_p = 1$ is the third component of the nucleon isospin, $q_p = -q_e = -q_\mu = e$ and $q_n = 0$ are charges, $m^*_{\mathrm{b}} \equiv m_{\mathrm{b}} + \textsl{g}_{\sigma \mathrm{b}}\sigma$ is the effective nucleon mass, $g_{\xi\mathrm{b}}$ with $\xi=\sigma,\omega,\rho$ the coupling constants, and $n = \sum_{\mathrm{b}} n_{\mathrm{b}}$ the baryon number density. Due to time-reversal symmetry, the boson fields $\sigma$, $\omega$, $\rho$ and $A$ only take the temporal parts, and $\omega_{\mu\nu}$, $\rho_{\mu\nu}$, $A_{\mu\nu}$ represent their field tensors.

In the density-dependent RMF models, the baryon-meson coupling constants are allowed to vary with the baryon density $n$, rather than being constant over the entire density range~\cite{Typel1999_NPA656-331}. The coupling constants $g_{\xi\mathrm{b}}$ ($\xi=\sigma$ and $\omega$) are thus density-dependent and take the following form, i.e.,
\begin{equation}
g_{\xi\mathrm{b}}(n) = g_{\xi\mathrm{b}}(n_0) a_{\xi} \frac{1+b_{\xi}(n/n_0+d_{\xi})^2}
                          {1+c_{\xi}(n/n_0+e_{\xi})^2}, \label{eq:ddcp_TW}
\end{equation}
where $n_0$ is the nuclear saturation density. For the $\boldsymbol{\rho}$ meson, the following form is used:
\begin{equation}
g_{\rho\mathrm{b}}(n) = g_{\rho\mathrm{b}}(n_0) \exp{\left[-a_\rho(n/n_0-1)\right]}. \label{eq:ddcp_rho}
\end{equation}

\begin{table}[h]
\centering
\caption{\label{Tab:Table 1}Saturation properties of nuclear matter predicted by the RDFs TW99~\cite{Typel1999_NPA656-331}, PKDD~\cite{Long2004_PRC69-034319} and DD-ME2~\cite{Lalazissis2005_PRC71-024312}. Here $n_0$ represents the saturation density of nuclear matter, $B$ the binding energy, $K$ the incompressibility, $J$ the skewness, $S$ the symmetry energy, $L$ and $K_\mathrm{sym}$ the slope and curvature of  symmetry energy.}
\begin{tabular}{l|ccccccc} \hline \hline
$ $ & $n_0$ & $B$ & $K$ & $J$ & $S$ & $L$ & $K_\mathrm{sym}$ \\
 & fm${}^{-3}$ & MeV & MeV & MeV & MeV & MeV & MeV    \\ \hline
TW99   & 0.153 & $-16.24$  & 240.2 & $-540 $ & 32.8 & 55.3  & $-125$ \\
PKDD   & 0.150 & $-16.27$  & 262.2 & $-119 $ & 36.8 & 90.2  & $-81 $\\
DD-ME2 & 0.152 & $-16.13$  & 250.8 & $477  $ & 32.3 & 51.2  & $-87 $\\
\hline
\end{tabular}
\end{table}

The saturation properties of nuclear matter predicted by the RDFs TW99~\cite{Typel1999_NPA656-331}, PKDD~\cite{Long2004_PRC69-034319} and DD-ME2~\cite{Lalazissis2005_PRC71-024312} are listed in Table~\ref{Tab:Table 1}. According to our previous investigation~\cite{Niu2026_PRC113-025804}, the nuclear EOSs predicted by TW99~\cite{Typel1999_NPA656-331} are generally consistent with various state-of-the-art constraints from chiral effective field theory, heavy-ion collision experiments, and multimessenger neutron star observations~\cite{Huth2022_Nature606-276}, which are nonetheless marginally consistent with the recent mass-radius measurements of PSR J0437-4715~\cite{Choudhury2024_ApJ971-L20} and PSR J0614-3329~\cite{Mauviard2025_TAJ995-60} as will be illustrated in Fig.~\ref{Fig:T2}. Compared with the density functional TW99~\cite{Typel1999_NPA656-331}, the functional PKDD~\cite{Long2004_PRC69-034319} predicts large slope of symmetry energy $L$, while that of DD-ME2~\cite{Lalazissis2005_PRC71-024312} predicts large skewness $J$ for symmetric nuclear matter, leading to too stiff EOSs and too large radii for neutron stars. The inclusion of new degree of freedom would thus soften the EOSs predicted by PKDD and DD-ME2 so that they can be more compatible with various state-of-the-art constraints.

\begin{table}[h]
\centering
\caption{\label{Tab:Table 2}Strangeness $S$, mass $m$, third component of isospin $\tau_{3}$, spin and parity $J^{P}$,
charge $q$, and coupling constants $\alpha_{\xi\mathrm{b}}=g_{\xi\mathrm{b}}/g_{\xi N}$ ($\xi=\sigma$ and $\rho$) for $\Lambda^0$, $\Xi^{0,-}$,
and $\Sigma^{+,0,-}$ hyperons and $\Delta$ baryons. Note that $\alpha_{\omega\mathrm{b}}=1$ is adopted in this work.}
\begin{tabular}{l|ccccc|cc|c}
\hline
              &      &               &            &             &               & PKDD           & DD-ME2                & All \\
      b       & $S$  & $m$ (MeV)     & $\tau_{3}$ & $J^{p}$     & $q$ ($e$)     & $\alpha_{\sigma\mathrm{b}}$& $\alpha_{\sigma\mathrm{b}}$       & $\alpha_{\rho\mathrm{b}}$\\ \hline
 $\Lambda^0$  & $-1$ & $1115.6$      &  $0$       & $(1/2)^{+}$ & $0$           & 0.878          &  0.877                &  0            \\ \hline
 $\Xi^0$      & $-2$ & $1314.9$      &  $-1$      & $(1/2)^{+}$ & $0$           & 0.845          &  0.844                &  1            \\
 $\Xi^-$      & $-2$ & $1321.3$      &  $+1$      & $(1/2)^{+}$ & $-1$          & 0.845          &  0.844                &  1            \\ \hline
 $\Sigma^+$   & $-1$ & $1189.4$      &  $-2$      & $(1/2)^{+}$ & $+1$          & 0.730          &  0.728                &  1            \\
 $\Sigma^0$   & $-1$ & $1192.5$      &  $0$       & $(1/2)^{+}$ & $0$           & 0.730          &  0.728                &  1            \\
 $\Sigma^-$   & $-1$ & $1197.4$      &  $+2$      & $(1/2)^{+}$ & $-1$          & 0.730          &  0.728                &  1            \\ \hline
$\Delta^{++}$ & $0$  & $1112$,$1232$ &  $-3$      & $(3/2)^{+}$ & $+2$          & 1              &     1                 &  0,1            \\
 $\Delta^{+}$ & $0$  & $1112$,$1232$ &  $-1$      & $(3/2)^{+}$ & $+1$          & 1              &     1                 &  0,1            \\
 $\Delta^0$   & $0$  & $1112$,$1232$ &  $0$       & $(3/2)^{+}$ & $0$           & 1              &     1                 &  0,1            \\
 $\Delta^-$   & $0$  & $1112$,$1232$ &  $+3$      & $(3/2)^{+}$ & $-1$          & 1              &     1                 &  0,1            \\
\hline\hline
\end{tabular}
\end{table}

For the $\omega$-hyperon couplings, following our previous investigations~\cite{Sun2018_CPC42-025101, Sun2019_PRD99-023004}, we take $g_{\omega \Lambda} = g_{\omega \Xi} = g_{\omega \Sigma} = g_{\omega N}$, which implies that the $\phi$-baryon coupling vanishes according to SU(3) symmetry-inspired prescriptions for meson-baryon couplings~\cite{Schaffner1996_PRC53-1416, Weissenborn2012_PRC85-065802, Miyatsu2013_PRC88-015802, Oertel2015_JPG42-075202}. Therefore, we neglect the contributions of $\sigma^*$ and $\phi$ mesons in the Lagrangian density in Eq.~(\ref{eq:Lagrange_1}). The $\sigma$-hyperon couplings are then set according to their potential depths in symmetric nuclear matter at saturation density $n_0$, i.e., the $\Lambda$ potential depth $V_\Lambda = - 29.786$~MeV, the $\Xi$ potential depth $V_\Xi = -16.276$~MeV, and the $\Sigma$ potential depth $V_\Sigma = 30$~MeV are adopted, which well reproduce the experimental single $\Lambda$ and $\Xi$ energy levels in finite nuclei~\cite{Liu2018_PRC98-024316}. The $\sigma$-hyperon couplings for the three models are listed in Table~\ref{Tab:Table 2}. For the $\rho$-hyperon couplings, we adopt $g_{\rho \Lambda} = 0$ and $g_{\rho \Xi} = g_{\rho \Sigma} = g_{\rho N}$.

For the $\omega$-$\Delta$ and $\sigma$-$\Delta$ couplings, since the potential depth of $\Delta$ in nuclear medium is similar to that of nucleons, we take $g_{\omega \Delta} = g_{\omega N}$ and $g_{\sigma \Delta} = g_{\sigma N}$~\cite{Li1997_PRC56-1570, Kosov1998_PLB421-37, Drago2014_PRC90-065809, Drago2014_PRD89-043014, Zhu2016_PRC94-045803}. In contrast, limited information is available for the $\rho$-$\Delta$ coupling. We thus consider two scenarios, i.e., the universal baryon-meson coupling scheme $g_{\rho \Delta}= g_{\rho N}$ and $g_{\rho \Delta}= 0$. Since the $\Delta$ isobar has a Breit-Wigner mass distribution around $1232$~MeV with a width of $\sim$$120$~MeV, we take $m_\Delta = 1112$~MeV and $m_\Delta = 1232$~MeV, while larger $m_\Delta$ ($\approx 1352$ MeV) would have little impact on neutron star properties~\cite{Sun2019_PRD99-023004}. The properties and coupling constants of all non-nucleonic baryons appearing in Eq.~(\ref{eq:Lagrange_1}) are summarized in Table~\ref{Tab:Table 2}.

From the Lagrangian density in Eq.~(\ref{eq:Lagrange_1}), the meson fields can be obtained as
\begin{eqnarray}
m_\sigma^2 \sigma &=& -\sum_\mathrm{b} g_{\sigma\mathrm{b}} n_\mathrm{b}^\mathrm{s}, \label{eq:eom_sigma} \nonumber\\
m_\omega^2 \omega_0 &=& \sum_\mathrm{b} g_{\omega\mathrm{b}} n_\mathrm{b},  \label{eq:eom_omega}\\
m_\rho^2 \rho_{0,3} &=& \sum_\mathrm{b} g_{\rho\mathrm{b}}\tau_{\mathrm{b},3} n_\mathrm{b}, \nonumber \label{eq:eom_rho}
\end{eqnarray}
where $n_\mathrm{b} = \langle \bar{\psi}_\mathrm{b}\gamma^0 \psi_\mathrm{b} \rangle$ and $n_\mathrm{b}^\mathrm{s} = \langle \bar{\psi}_\mathrm{b}\psi_\mathrm{b} \rangle$ are the number and scalar densities of baryon b, respectively. At zero temperature, the total energy density is given by
\begin{eqnarray}
\varepsilon &=& \sum_{i={\mathrm{b},l}}\varepsilon_i(\nu_i, m_i^*) + \sum_{\xi=\sigma, \omega, \boldsymbol{\rho}} \frac{1}{2}m_\xi^2 \xi^2,
\label{eq:E}
\end{eqnarray}
where the kinetic energy density of each fermion species $i$ adopting no-sea approximation is
\begin{eqnarray}
\varepsilon_i(\nu_i, m_i) &=& \int_0^{\nu_i} \frac{f_i p^2}{2\pi^2} \sqrt{p^2+m_i^2}\mbox{d}p \label{eq:ei0}\\
&=&  \frac{f_i m_i^4}{16\pi^{2}} \left[x_i(2x_i^2+1)\sqrt{x_i^2+1}-\mathrm{arcsh}(x_i) \right].
\nonumber
\end{eqnarray}
Here $x_i \equiv \nu_i / m_i$ with $\nu_i$ being the Fermi momentum and $f_i = 2 J_i + 1$ the degeneracy factor of particle species $i$. In Eq.~(\ref{eq:E}), the effective baryon mass is $m_\mathrm{b}^* \equiv m_\mathrm{b} + g_{\sigma\mathrm{b}} \sigma$, while the lepton masses remain constant, i.e., $m_l^* \equiv m_l$. The number and scalar densities are expressed as
\begin{eqnarray}
n_i &=& \langle \bar{\psi}_i\gamma^0 \psi_i \rangle = \frac{f_i\nu_i^3}{6\pi^2}, \label{eq:ni} \\
n_i^\mathrm{s}&=&\langle \bar{\psi}_i\psi_i \rangle
=\frac{f_i m_i^3}{4\pi^2} \left[x_i\sqrt{x_i^2+1} - \mathrm{arcsh}(x_i)\right]. \label{eq:ns}
\end{eqnarray}

The chemical potentials of baryons and leptons are then given by
\begin{eqnarray}
\mu_\mathrm{b}&=& g_{\omega\mathrm{b}} \omega_0
              + g_{\rho\mathrm{b}}\tau_{\mathrm{b},3} \rho_{0,3}
              + \Sigma^\mathrm{R}
              + \sqrt{\nu_\mathrm{b}^2+{m_\mathrm{b}^*}^2},
\label{eq:chem_B} \\
\mu_l &=&  \sqrt{\nu_l^2+m_l^2},
\label{eq:chem_l}
\end{eqnarray}
where the rearrangement term is
\begin{eqnarray}
&&\Sigma^\mathrm{R}= \sum_{\mathrm{b}}\left(
   g_{\sigma \mathrm{b}}' \sigma n_\mathrm{b}^\mathrm{s}+
   g_{\omega\mathrm{b}}' \omega_0 n_\mathrm{b}+
   g_{\rho\mathrm{b}}' \rho_{0,3}\tau_{\mathrm{b},3} n_\mathrm{b}
  \right).
\label{eq:re_B}
\end{eqnarray}
The pressure is given by
\begin{equation}
p = \sum_i \mu_i n_i \nonumber - \varepsilon. \label{eq:pressure}
\end{equation}
For neutron star matter, the charge neutrality condition must be satisfied:
\begin{equation}
  \sum_i q_i n_i = 0, \label{eq:Chntr}
\end{equation}
where $q_i$ denotes the charge of species $i$. To achieve the minimum energy configuration, particles undergo weak interactions until the $\beta$-equilibrium condition is established, i.e.,
\begin{equation}
\mu_\mathrm{b}= \mu_n - q_\mathrm{b} \mu_e,~~ \mu_\mu = \mu_e.  \label{eq:weakequi}
\end{equation}
It should be noted that in all three RDFs considered, the inner crust region of the neutron star is included, where nuclear matter becomes nonuniform. The theoretical details of this part have been discussed in our previous work~\cite{Sun2025_PRD111-103019}; see also Ref.~\cite{Xia2022_CTP74-095303} for further theoretical formulations, which will not be repeated here.

\begin{figure}
\includegraphics[width=\linewidth]{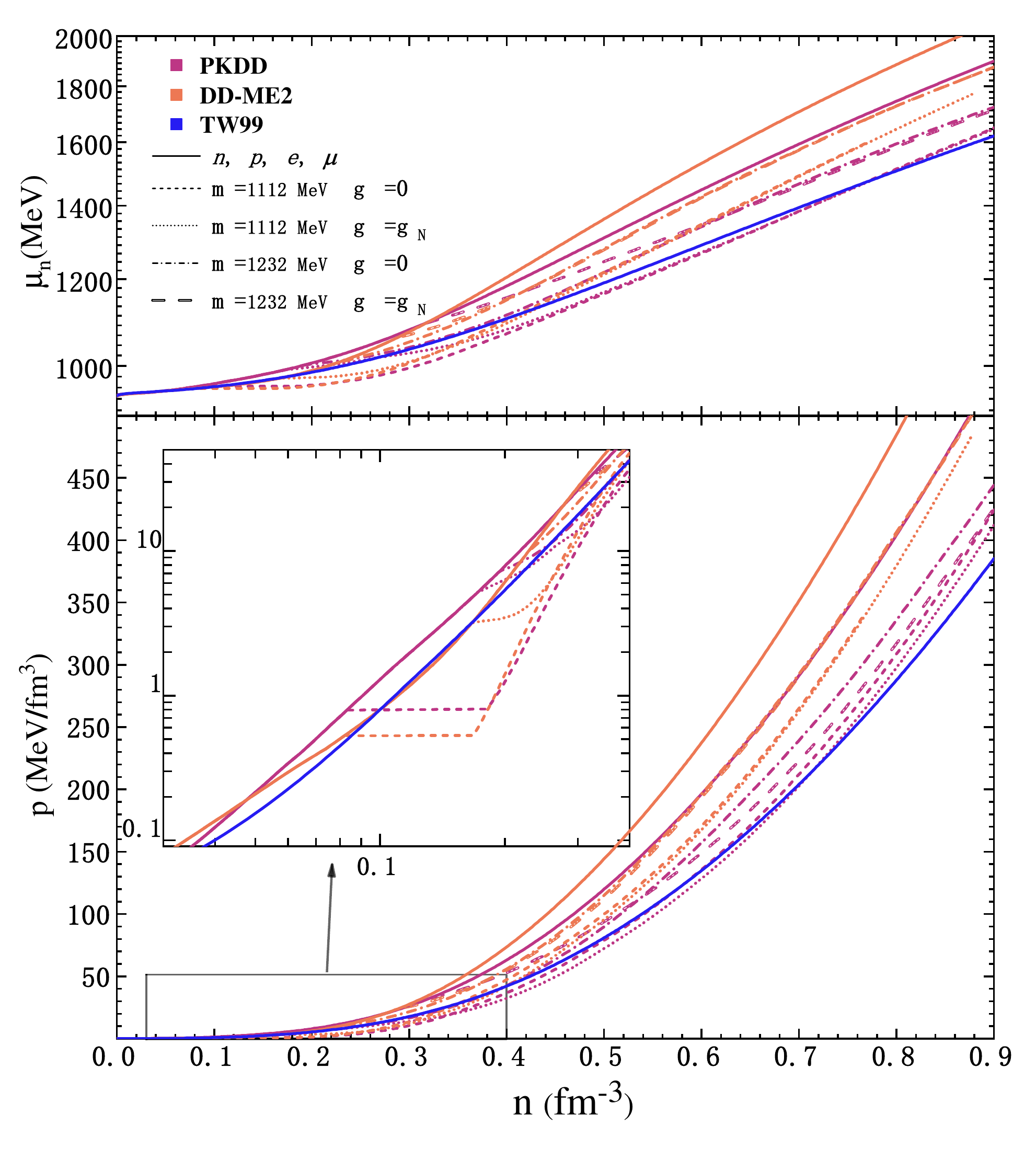}
\caption{\label{Fig:T1} Nucleon chemical potential ($\mu_n$) and pressure ($p$) of neutron star matter as functions of baryon number density $n$, which are obtained adopting three nuclear RDFs (TW99, PKDD, and DD-ME2), various baryons and meson-baryon couplings. In particular, the solid lines represent the ordinary $npe\mu$ matter, while the various types of dashed lines represent the $\Delta$ admixed matter with different masses ($m_\Delta = 1112$~MeV and $m_\Delta = 1232$~MeV) and coupling coefficients ($g_{\rho \Delta}= g_{\rho N}$ and $g_{\rho \Delta}= 0$) listed in Table~\ref{Tab:Table 2}.}
\end{figure}

In Fig.~\ref{Fig:T1} we present the nucleon chemical potential ($\mu_n$) and pressure ($p$) of neutron star matter as functions of baryon number density $n$, which are predicted adopting three nuclear RDFs (TW99, PKDD, and DD-ME2), various baryons and meson-baryon couplings listed in Table~\ref{Tab:Table 2}. Various dashed curves denote the EOSs incorporating $\Delta$ resonances with different $\Delta$ baryon masses ($m_\Delta = 1112$~MeV and $m_\Delta = 1232$~MeV) and $\rho$-$\Delta$ couplings ($g_{\rho\Delta} = g_{\rho N}$ and 0). It should be noted that the sub-saturation density portion of the EOSs at $n \lesssim 0.08 \text{ fm}^{-3}$ corresponds to neutron stars' crusts, where we have adopted the results presented in Refs.~\cite{Xia2022_CTP74-095303, Niu2026_PRC113-025804, Xia2021_PRC103-055812}. Therefore, the crustal EOSs are identical if the same nuclear RDF is adopted. As expected, the EOS becomes softer once new degrees of freedom emerge. The smaller $m_\Delta$ and $g_{\rho\Delta}$, the earlier the $\Delta$ baryons affect the EOSs as density increases. Particularly, for the cases of $m_\Delta = 1112 \text{ MeV}$ and $g_{\rho\Delta} = 0$, a first-order phase transition from nuclear matter to $\Delta$-admixed matter occurs in the density range of $n = 0.083$-$0.17 \text{ fm}^{-3}$, which will significantly impact the intrinsic properties of neutron stars.

\begin{figure*}
\includegraphics[width=\linewidth]{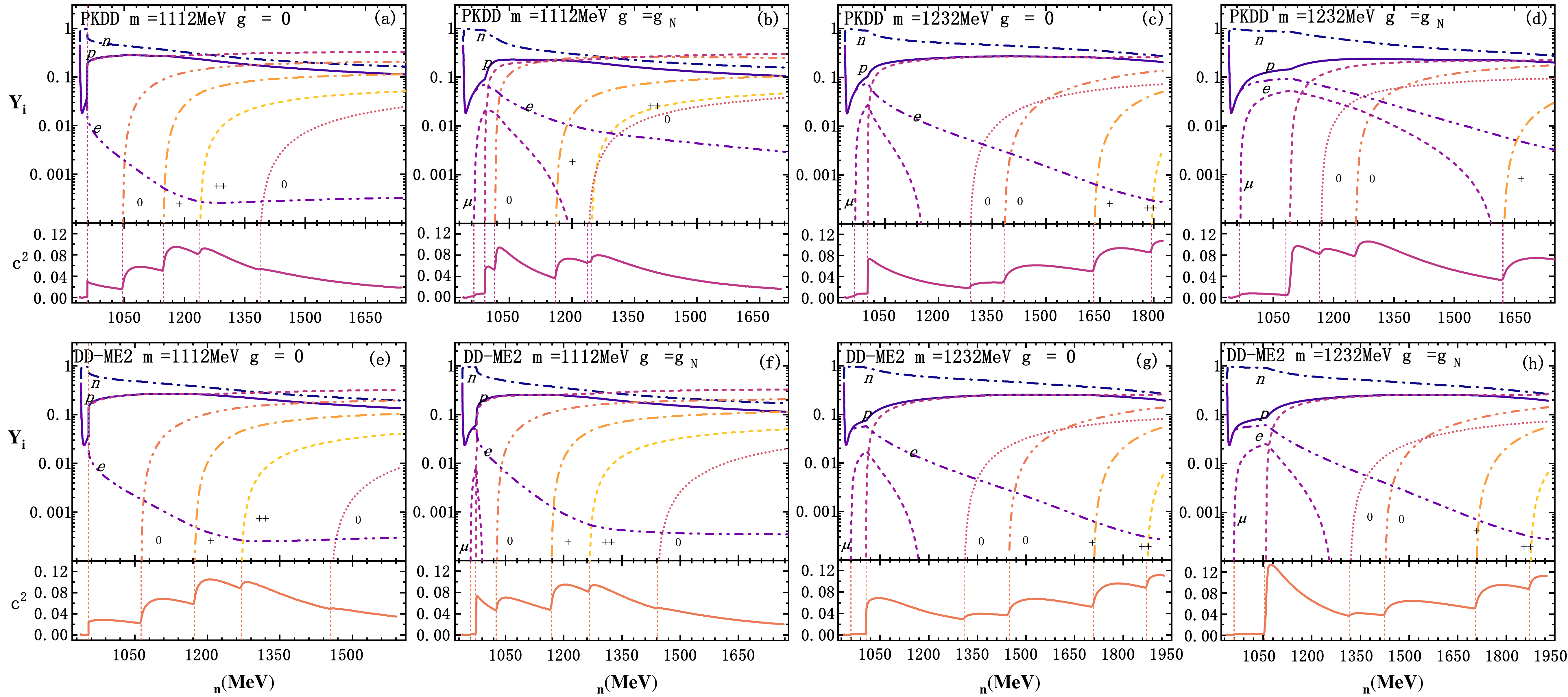}
\caption{\label{Fig:T3}Particle fractions ($Y_i = n_i / n$) and difference of the squared sound speeds ($\Delta c^2=c^2_s-c^2_e$) as functions of neutron chemical potential ($\mu_n$), where the nuclear RDFs PKDD and DD-ME2 are adopted. The first row displays the results for the PKDD model, where subplots (a), (b), (c), and (d) represent the $\Delta$ resonance parameters ($m_\Delta = 1112 \text{ MeV}$, $g_{\rho\Delta} = 0$), ($m_\Delta = 1112 \text{ MeV}$, $g_{\rho\Delta} = g_{\rho N}$), ($m_\Delta = 1232 \text{ MeV}$, $g_{\rho\Delta} = 0$), and ($m_\Delta = 1232 \text{ MeV}$, $g_{\rho\Delta} = g_{\rho N}$), respectively. The second row shows the corresponding subplots (e), (f), (g), and (h) for the DD-ME2 model. The vertical dashed lines in the $\Delta c^2$ subplots indicate the onset of new particles, and the curves for each particle species are explicitly labeled in the $Y_i$ panels.}
\end{figure*}

Figure~\ref{Fig:T3} illustrates the relationship of the particle fractions  $Y_i = n_i / n$ ($i = n, p, e, \mu$, hyperons, $\Delta$ resonances) and the difference of the squared sound speeds ($\Delta c^2=c^2_s-c^2_e$) with respect to neutron chemical potential ($\mu_n$) adopting nuclear RDFs PKDD and DD-ME2. The equilibrium sound speed $c_e$ and adiabatic sound speed $c_s$ are fixed by~\cite{Jaikumar2021_PRD103-123009, Zheng2023_PRD107-103048}
\begin{equation}
    c^2_e =  \frac{\mbox{d}p}{\mbox{d}\varepsilon}, \ \ \ c^2_s = \left.\frac{\mbox{d}p}{\mbox{d}\varepsilon}\right|_{\{Y_i\}},    \label{eq:14}
\end{equation}
whose difference determines the local $g$-mode frequency, i.e., Brunt-V\"{a}is\"{a}l\"{a} frequency as indicated in Eq.~(\ref{eq:13}). The difference of the sound speeds arises due to the emergence of new degrees of freedom, so that the $g$-mode oscillations in neutron stars provide crucial tool to probe their internal composition. We present all combinations of $\Delta$ baryon masses ($m_\Delta = 1232 \text{ MeV}$, $m_\Delta = 1112 \text{ MeV}$) and $\Delta$-$\rho$ coupling constants ($g_{\rho\Delta} = g_{\rho N}$, $g_{\rho\Delta} = 0$). As $\mu_n$ surpasses the rest mass of neutrons, they will drip out of nuclei and form a neutron gas, leading to nonzero $\Delta c^2$~\cite{Niu2026_PRC113-025804, Sun2025_PRD111-103019}. The proton fraction $Y_p$ in neutron star matter thus decreases with $\mu_n$ until reaching its minimum at $\mu_n \approx 945$ MeV. Then muons emerge at $\mu_n \approx 960$ MeV with $\Delta c^2$ increases slightly, while $Y_p$ increases drastically due to the neutralizing effects of muons.

According to Fig.~\ref{Fig:T3}, we can observe that among all $\Delta$ baryons, the negatively charged $\Delta^-$ appears first as $\mu_n$ increases. This is because the charge neutrality condition suppresses the emergence of the positively charged $\Delta^+$. The electron and muon fractions decrease rapidly with the onset of $\Delta^-$, while the $\Delta$ onset density increases with both $m_\Delta$ and $g_{\rho\Delta}$, in consistent with our previous study~\cite{Sun2019_PRD99-023004}. When we adopt smaller values for $m_\Delta$ and $g_{\rho\Delta}$, the impact of $\Delta$ resonances becomes more significant, with $\Delta^-$, $\Delta^0$, $\Delta^+$, and $\Delta^{++}$ appearing sequentially as the density increases. In the case of $m_\Delta = 1112 \text{ MeV}$ and $g_{\rho\Delta} = 0$, the hyperon $\Lambda^0$ appears latest, while the emergence of other hyperons (such as $\Sigma^-$, $\Xi^0$, etc.) is further hindered and emerges at much higher chemical potential beyond the scope of neutron stars. Note that in this case there exist discontinuities in $Y_i$ and $\Delta c^2$ around $\mu_n \approx 956 \text{ MeV }$ in panels (a) and (e), which are due to a first-order phase transition from nuclear matter to $\Delta$-admixed matter. Comparing the $\Delta c^2$ panels below, it is clearly observed that the difference in sound speeds increases sharply whenever a new degree of freedom appears. This is primarily due to the sudden drop in $c_e^2$, as discussed in Refs.~\cite{Tran2023_PRC108-015803, Jaikumar2021_PRD103-123009, Wei2020_ApJ904-187, Constantinou2021_PRD104-123032, Zhao2022_PRD105-103025, Zheng2023_PRD107-103048}. This effect is particularly pronounced with the onset of $\Delta$ resonances (especially $\Delta^-$). Compared to pure $npe\mu$ matter, the $\Delta$ resonances have a particularly significant impact on the difference in squared sound speeds because the fraction of $\Delta^-$ is relatively large, while the contribution from leptons is relatively small.

\subsection{Neutron star structure}

Due to the strong gravitational field inside neutron stars, their structures and evolutions are governed by Einstein's field equations in general relativity, i.e.,
\begin{equation}
R_{\nu\mu} - \frac{1}{2} g_{\nu\mu}R = 8\pi T_{\nu\mu}, \label{eq:1}
\end{equation}
where $R_{\mu\nu}$ is the Ricci tensor and $R$ the Ricci scalar. The line element for a static and spherically symmetric spacetime is given by the Schwarzschild metric, i.e.,
\begin{equation}
\mbox{d}s^{2} =-\mathrm{e}^{2\Phi } \mbox{d}t^{2} +\mathrm{e}^{2\Lambda}\mbox{d}r^2+r^2(\mbox{d}\theta^{2}+\sin^2\theta \mbox{d}\phi ^2 ),
\label{eq:2}
\end{equation}
where $\Phi$ and $\Lambda$ are metric functions depending on $r$. For a perfect fluid, the energy-momentum tensor takes the form
\begin{equation}
T_{\mu \nu} = (\varepsilon + p)u_{\mu }u_{\nu } + p g_{\mu \nu}. \label{eq:3}
\end{equation}
The mass function $m(r) = r(1 - e^{-2\Lambda})/2$ satisfies
\begin{equation}
\frac{\mathrm{d}m}{\mathrm{d}r} = 4\pi r^2 \varepsilon, \label{eq:4}
\end{equation}
where $\varepsilon$ is the energy density and $p$ the pressure. To determine the radial profiles of $p(r)$ and $\Phi(r)$, one needs to solve the Tolman-Oppenheimer-Volkov (TOV) equations, i.e.,
\begin{eqnarray}
 \frac{\mathrm{d} p}{\mathrm{~d} r}&=&-(\varepsilon+p) \frac{\mathrm{d} \Phi}{\mathrm{d} r}, \label{eq:TOV5}\\
 \frac{\mathrm{d} \Phi}{\mathrm{d} r}&=&\frac{m+4 \pi r^{3} p}{r(r-2 m)}. \label{eq:TOV6}
\end{eqnarray}

\begin{figure}
\includegraphics[width=\linewidth]{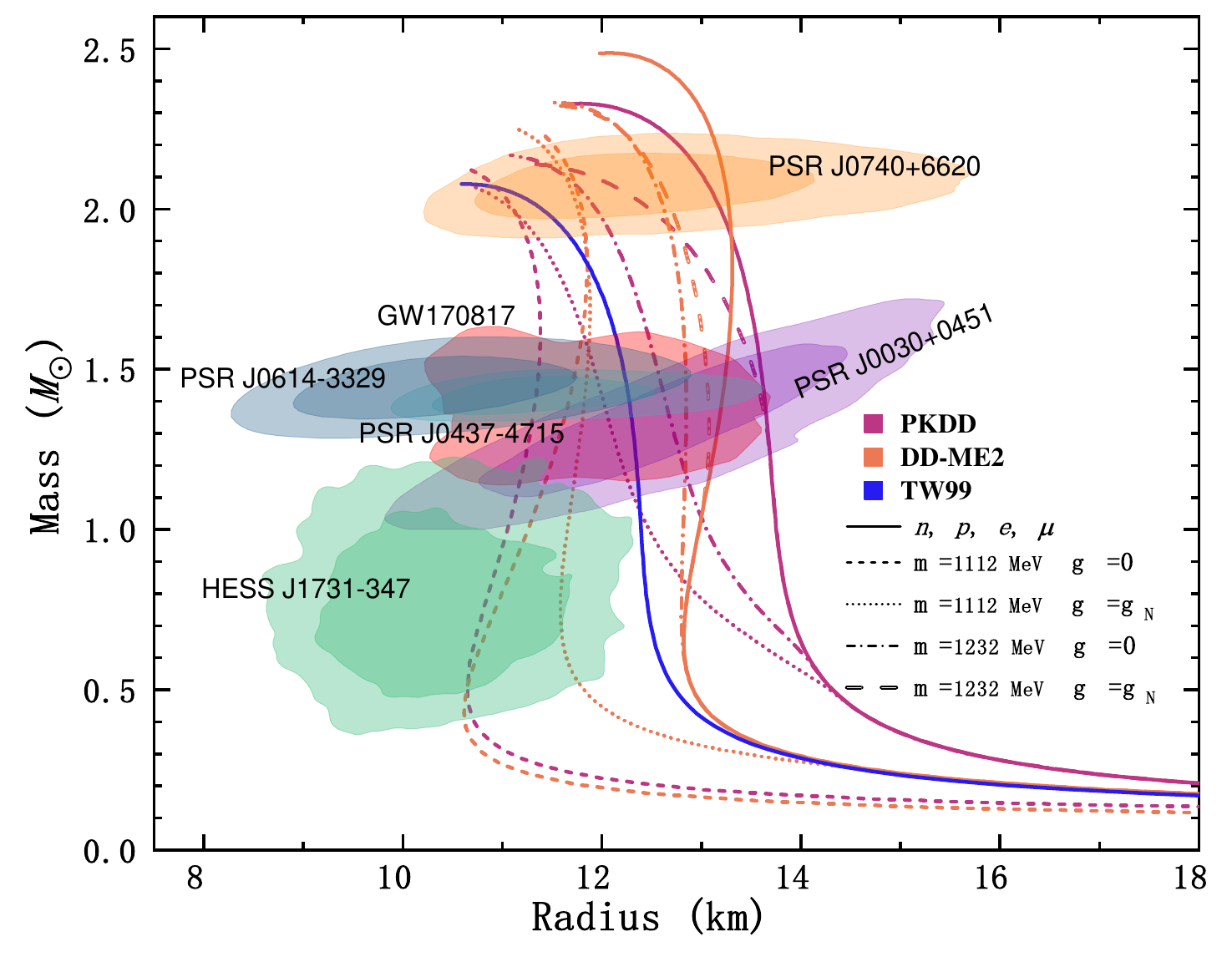}
\caption{\label{Fig:T2}Mass-radius relations of neutron stars predicted by the EOSs presented in Fig.~\ref{Fig:T1}, where three nuclear RDFs (TW99, PKDD, and DD-ME2), various baryons and meson-baryon couplings are adopted. The solid curves represent traditional neutron stars, while various types of dashed lines represent neutron stars with hyperons and $\Delta$ resonances with different $m_\Delta$ and $\Delta$-$\rho$ coupling. The shaded regions represent the observational masses and radii from the binary neutron star merger event GW170817 in 90\% confidence interval~\cite{LVC2018_PRL121-161101}; PSR J0030+0451, PSR J0740+6620, PSR J0437-4715, PSR J0614-3329, and HESS J1731-347 in the 68\% and 95\% confidence intervals~\cite{Riley2019_ApJL887-L21, Riley2021_ApJL918-L27, Miller2019_ApJL887-L24, Miller2021_ApJL918-L28, Choudhury2024_ApJ971-L20, Doroshenko2022_NatAs6_1444, Mauviard2025_TAJ995-60}.}
\end{figure}

Employing the EOSs indicated in Fig.~\ref{Fig:T1}, we then fix the structures of neutron stars by solving the TOV equations, where in Fig.~\ref{Fig:T2} we present the obtained mass-radius relations. Due to the softening in the EOSs, the maximum mass of neutron stars containing $\Delta$ resonances is reduced compared with those made of $npe\mu$ matter alone, while the radii of neutron stars become smaller with the emergence of $\Delta$ resonances as well. Compared with various astrophysical constraints on the masses and radii of neutron stars~\cite{LVC2018_PRL121-161101, Riley2019_ApJL887-L21, Riley2021_ApJL918-L27, Miller2019_ApJL887-L24, Miller2021_ApJL918-L28, Choudhury2024_ApJ971-L20, Doroshenko2022_NatAs6_1444, Mauviard2025_TAJ995-60}, the radii of traditional neutron stars made of $npe\mu$ matter are generally too large, while those predicted by TW99 are marginally consistent with the mass and radius measurements of PSR J0437-4715~\cite{Choudhury2024_ApJ971-L20} and PSR J0614-3329~\cite{Mauviard2025_TAJ995-60} but not HESS J1731-347~\cite{Doroshenko2022_NatAs6_1444}. The inclusion of $\Delta$ resonances improve the situation significantly. In particular, neutron stars hosting a first-order phase transition from nuclear matter to $\Delta$-admixed matter ($m_\Delta = 1112 \text{ MeV}$ and $g_{\rho\Delta} = 0$) can well accommodate those observational constraints even for HESS J1731-347, indicating the crucial role of $\Delta$ resonances in reconciling the theoretical neutron star structures with various astrophysical observations.

\section{\label{sec:nonradial} Non-radial oscillations in neutron stars}

In this work, we investigate the oscillation frequencies of the $g$, $f$, and $p$ modes, which are obtained under the full general relativistic framework as well as the Cowling approximation. Similar to previous works, we linearize Einstein's equation (\ref{eq:1}) to fix the differential equations governing small perturbations around equilibrium. If we further assume a static spacetime and allow only fluid oscillations within the neutron star, the differential equations can be significantly simplified, i.e., the well-known Cowling approximation. For clarity and completeness, we briefly restate some key definitions here, while more detailed derivations can be found in Refs.~\cite{Detweiler1985_ApJ292-12, Lindblom1983_ApJS53-73, Zhao2022_PRD105-103025}.

\subsection{Non-radial oscillation equations in full general relativity}

In the fully relativistic case, a small metric perturbation $h_{\mu\nu}$ is introduced on the static, spherically symmetric background. In the Regge-Wheeler gauge~\cite{Regge1957_PR108-1063}, the perturbed metric is written as
\begin{equation} \label{eq:pertmetric}
h_{\mu\nu} = \begin{pmatrix}
e^{2\Phi}r^{l}H_0 & r^{l+1}\dot{H} _1 & 0 & 0 \\
r^{l+1}\dot{H} _1 & e^{2\Lambda}r^{l}H_2 & 0 & 0 \\
0 & 0 & r^{l+2}K & 0 \\
0 & 0 & 0 & r^{l+2}\sin^2\theta K
\end{pmatrix} e^{i\omega t} Y^l_m,
\end{equation}
where $H_0$, $H_1$, $H_2$, and $K$ are metric perturbation functions, and $Y^l_m(\theta,\phi)$ the spherical harmonics with angular degree $l$ and azimuthal order $m$. Here $\omega$ is the complex oscillation frequency, whose real part represents the oscillation frequency and imaginary part corresponds to the inverse damping time of gravitational radiation.

Meanwhile, the fluid displacement vector inside the star can be expanded in spherical harmonics. For a non-rotating, zero-temperature neutron star, the Lagrangian displacement vector is given by
\begin{eqnarray}
    \xi^{r}&=&r^{l-2}\mathrm{e}^{-\Lambda}W Y^l_m \mathrm{e}^{i\omega t},     \label{eq:8}\\
    \xi^{\theta}&=&-r^{l-2}V\partial_{\theta} Y^l_m \mathrm{e}^{i\omega t},     \label{eq:9}\\
    \xi^{\phi}&=&-r^{l-2}(\sin\theta)^{-2}V\partial_{\phi} Y^l_m \mathrm{e}^{i\omega t},     \label{eq:10}
\end{eqnarray}
where $W(r)$ and $V(r)$ are eigenfunctions associated with the radial and tangential displacements, respectively. The fluid perturbation amplitude $X$ related to the Lagrangian pressure variation is defined as
\begin{equation} \label{eq:Xdef}
\Delta p=-r^l e^{-\Phi}X Y^l_m \mathrm{e}^{i\omega t}.
\end{equation}

The perturbation equations follow from the linearized Einstein equations together with energy-momentum conservation, where the oscillation equations can be reduced to a system of four coupled first-order differential equations~\cite{Detweiler1985_ApJ292-12}, i.e.,
\begin{eqnarray} \label{eq:osc1}
\frac{\mathrm{d} H_1}{\mathrm{d} r} &=&
-\frac{1}{r}\left[
l + 1 + \frac{2m e^{2\Lambda}}{r} + 4\pi r^2 e^{2\Lambda} (p - \varepsilon)
\right]H_1
 \nonumber \\
 &&\mbox{}
+ \frac{e^\lambda}{r}\left[
H_0 + K - 16\pi (\varepsilon + p)V
\right],
\\
\frac{\mathrm{d} K}{\mathrm{d} r} &=&
\frac{H_0}{r} + \frac{l(l+1)}{2r}H_1
- \left(\frac{l+1}{r} - \frac{\mathrm{d} \Phi}{\mathrm{d} r}\right)K
 \nonumber \\
 &&\mbox{}
- \frac{8\pi (\varepsilon + p)e^{\Lambda}}{r}W,
\\
\frac{\mathrm{d} W}{\mathrm{d} r} &=& -\frac{l+1}{r}W + r e^{\Lambda}\Bigg[\frac{X}{\Gamma_1 p e^{\Phi}} - \frac{l(l+1)}{r^2}V
 \nonumber \\
 &&\mbox{}
+ \frac{1}{2}H_0 + K\Bigg],
\\
\frac{\mathrm{d} X}{\mathrm{d} r} &=&
-\frac{l}{r}X + (\varepsilon + p)e^{\Phi}\Bigg[
\frac{1}{2}\left(\frac{1}{r} - \frac{\mathrm{d} \Phi}{\mathrm{d} r}\right)H_0
 \nonumber \\
 &&\mbox{}
 + \frac{1}{2}\left(\frac{r\omega^2}{e^{2\Phi}} + \frac{l(l+1)}{2r}\right)H_1
 \nonumber \\
 &&\mbox{}
 + \frac{1}{2}\left(3\frac{\mathrm{d} \Phi}{\mathrm{d} r} - \frac{1}{r}\right)K - \frac{l(l+1)}{r^2}\frac{\mathrm{d} \Phi}{\mathrm{d} r}V
 \nonumber \\
 &&\mbox{}
 - \frac{1}{r}\left[4\pi (\varepsilon + p)e^{\Lambda} + \frac{\omega^2 e^{\Lambda}}{e^{2\Phi}} \right.
  \nonumber \\
 &&\mbox{}
  \left.
 + \frac{1}{e^{\Lambda}}\left(\frac{\mathrm{d} \Phi}{\mathrm{d} r} \left(\frac{\mathrm{d}\Lambda}{\mathrm{d} r} + \frac{2}{r}\right) - \frac{\mathrm{d}^2 \Phi}{\mathrm{d} r^2} \right) W \right].
\label{eq:Lagrange}
\end{eqnarray}
The two remaining perturbation functions $H_0$ and $V$ can be expressed in terms of $H_1$, $K$, $W$, and $X$:
\begin{eqnarray}
&&H_0\left(2m + nr + Q\right)  =
 8\pi r^3 e^{-\Phi} X
 \nonumber \\
 &&\mbox{}
 -\left[nQ - \omega^2r^3e^{-2(\Lambda+\Phi)}\right]H_1
\label{eq:Lagrange} \\
 &&\mbox{}
+ \left[nr - \omega^2 r^3 e^{-2\Phi} - e^{2\Lambda} r^{-1}Q(2m - r + Q)\right]K,
\nonumber\\
\nonumber\\
&&V = \left[\frac{X}{\varepsilon + p} - \frac{Q}{r^3}e^{\Phi+\Lambda}W
 - \frac{H_0}{2}e^{\Phi}
\right]\frac{e^{\Phi}}{\omega^2},
\label{eq:12}
\end{eqnarray}

where $n = \frac{1}{2}(l - 1)(l + 2)$, $Q = m + 4\pi r^3 p$, and $\Gamma_1 = {(\varepsilon + p)}c_s^2/{p}$ is the adiabatic index with $c_s$ being the adiabatic sound speed.

At the stellar center ($r=0$), the boundary conditions are
\begin{eqnarray}
W(0) &=& 1,
\\
X(0) &=& \left(\varepsilon_{0}+p_{0}\right) e^{\Phi_{0}} \Bigg\{\Bigg[\frac{4 \pi}{3}\left(\varepsilon_{0}+3 p_{0}\right)
\nonumber \\
 &&\mbox{}
-\frac{\omega^{2}}{l}e^{-2\Phi_{0}}\Bigg]W(0)+\frac{K(0)}{2}\Bigg\},
\\
H_1(0) &=& \frac{2l K(0) + 16\pi (\varepsilon_0 + p_0)W(0)}{l(l+1)},
\\
H_0(0) &=& K(0).
\label{eq:bc_center}
\end{eqnarray}
At the stellar surface ($r=R$), the function $X(R)$ must vanish, i.e., the Lagrangian perturbation of pressure is zero. The metric perturbations $H_1$, $H_0$, and $K$ must be matched continuously to the exterior solution.
In the vacuum exterior, fluid perturbations vanish and the system reduces to two ordinary differential equations for $H_1$ and $K$, which together form the Zerilli equation~\cite{Gittins2025_PRD111-083024}:
\begin{equation} \label{eq:zerilli}
\frac{d^2 Z}{dr_*^2} + \left[ \omega^2 - V_Z(r) \right] Z = 0
\end{equation}
where $V_Z(r)$ is the effective potential and $r_*$ is the tortoise coordinate. Gravitational waves generated by stellar oscillations must correspond to purely outgoing waves at infinity, where the eigenfrequencies $\omega$ are thus determined by enforcing this boundary condition.

\subsection{Cowling approximation}

Neglecting the spacetime perturbations induced by stellar oscillations and considering only the fluid perturbations leads to the Cowling approximation~\cite{Cowling1941_MNRAS101-367}. This simplification greatly reduces computational cost and has been verified by many studies~\cite{Zhao2022_PRD105-103025, Xu2017_PRD96-083005, Yoshida2002_AAP395-201, Ranea-Sandoval2018_JCAR12-031}, especially for $g$-modes where buoyancy acts as the main restoring force. As discussed in Ref.~\cite{Wei2020_ApJ904-187}, the frequency difference between the Cowling approximation and full GR calculations for $g$-modes is typically below 10\%, and even smaller for less massive neutron stars. However, the discrepancy for $f$-modes can be as large as 20\%~\cite{Zhao2022_PRD105-103025,Yoshida1997_MNRAS117-122-289,Sotani2020_PRD102-063025}.
By setting $H_0 = H_1 = K = 0$ in the full GR system, one obtains two equations involving only the fluid perturbation functions $W$ and $V$:
\begin{eqnarray}
    \frac{\mathrm{d} W}{\mathrm{d} r} &=& \frac{\mathrm{d}\varepsilon}{\mathrm{d} p}\left[\omega^2 r^2 \mathrm{e}^{\Lambda-2\Phi}V+\frac{\mathrm{d}\Phi}{\mathrm{d} r}W \right]-l(l+1)\mathrm{e}^{\Lambda}V,     \label{eq:37}\\
    \frac{\mathrm{d} V}{\mathrm{d} r} &=& \mathrm{e}^{\Lambda}\left(\frac{N^2}{\omega^2} - 1\right)\frac{W}{r^2}+\frac{N^2}{g \mathrm{e}^{2\Phi-2\Lambda}}V+2gV.     \label{eq:38}
\end{eqnarray}
Here $N$ is the Brunt-V\"{a}is\"{a}l\"{a} frequency~\cite{John1980_PUP}, representing the local oscillation frequency of fluid elements and serving as a local $g$-mode frequency:
\begin{equation}
    N^2 = g^2\left(\frac{1}{c^2_e}-\frac{1}{c^2_s}\right)\mathrm{e}^{2\Phi-2\Lambda}     \label{eq:13}
\end{equation}
where $g = - (\mbox{d}p/\mbox{d}r) / (\varepsilon + p)$. The difference between $1/c_e^2$ and $1/c_s^2$ with $c_e$ the equilibrium sound speed and $c_s$ the adiabatic sound speed determines the Brunt-V\"{a}is\"{a}l\"{a} frequency, and the global $g$-mode frequencies can be regarded as the averaged local buoyancy frequencies.

For a given EOS, the non-radial oscillation equations (\ref{eq:37}) and (\ref{eq:38}) together with the boundary conditions at the center and surface form a Sturm-Liouville eigenvalue problem. At the stellar center, we set
\begin{equation}
  W(0)+l V(0) = 0,     \label{eq:16}
\end{equation}
while at the surface ($r = R$) the Lagrangian pressure perturbation must vanish, i.e.,
\begin{equation}
    \omega^2 \mathrm{e}^{\Lambda-2\Phi}V(R)+\frac{1}{R^2}\frac{\mbox{d}\Phi}{\mbox{d}r}(R)W(R)=0.     \label{eq:40}
\end{equation}
By varying $\omega$ and solving for $V(r)$ and $W(r)$, the eigenfrequencies are found fulfilling the boundary conditions Eqs.~(\ref{eq:16}) and (\ref{eq:40}).

\subsection{Non-radial oscillations}

Here we focus on the quadrupole scenarios with $l = 2$. Adopting either of the two approaches described above, discrete eigenfrequencies can be obtained, whose values typically follow
\begin{equation}
\omega_{g_n} < \cdots < \omega_{g_1} < \omega_f < \omega_{p_1} < \cdots < \omega_{p_n}. \label{eq:18}
\end{equation}
Here $n$ denotes the number of radial nodes with $\omega_{g_n} \to 0$ and $\omega_{p_n} \to \infty$ as $n \to \infty$. The fundamental $f$-mode corresponds to $n = 0$, the $p$-modes are pressure modes with pressure being the restoring force, and the $g$-modes are gravity modes with buoyancy being the restoring force. The $g$-modes induced by chemical stratification are highly sensitive to the composition of the neutron-star core, making them potentially powerful probes of the internal structure.

\begin{figure*}
\includegraphics[width=\linewidth]{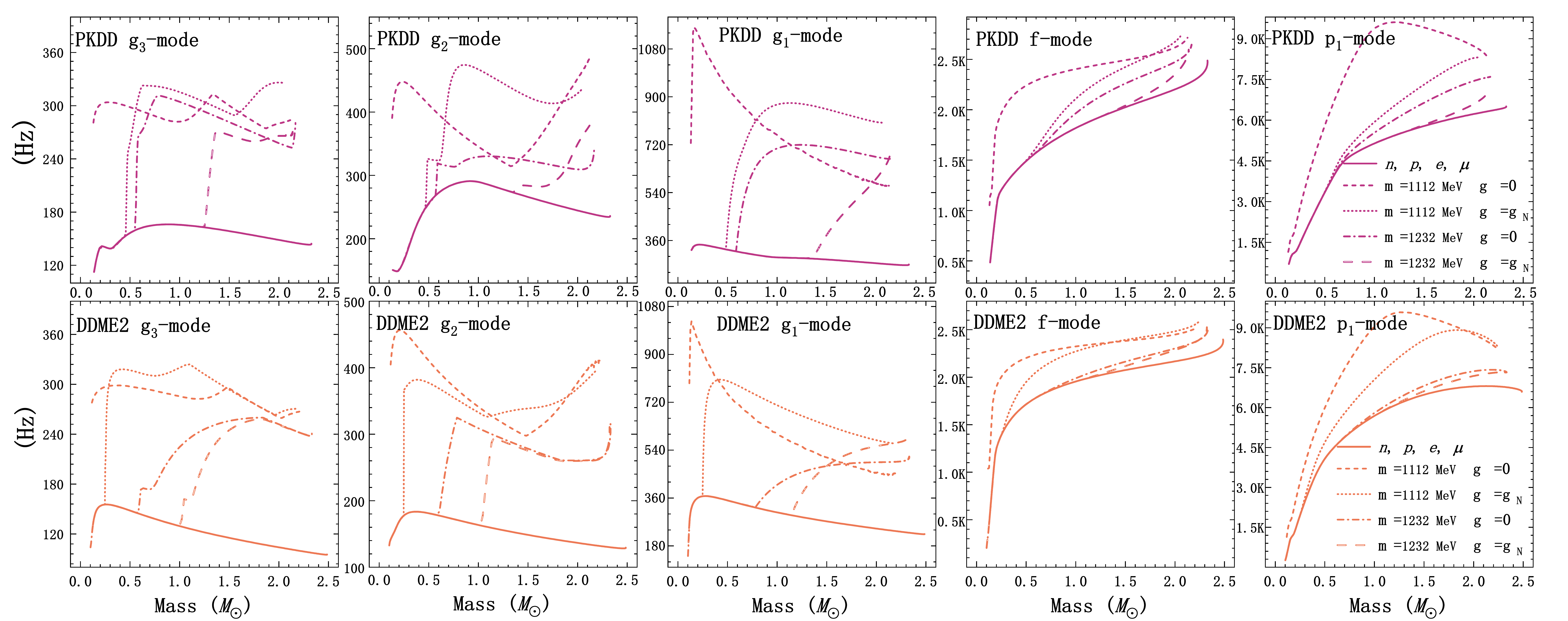}
\caption{\label{Fig:T5} Frequencies of $g$, $f$, and $p$-mode oscillations ($\nu$) as functions of neutron star mass $M$, which are predicted adopting the nuclear RDFs PKDD and DD-ME2, various baryons, and meson-baryon couplings. The solid curves represent traditional neutron stars, while those predicted by TW99 are expected to be similar and are thus not illustrated here.}
\end{figure*}

In Fig.~\ref{Fig:T5} we present the obtained $g$, $f$, and $p$-mode frequencies ($\nu$) as functions of neutron star mass $M$, which are obtained employing the Cowling approximation. Three overtones for the $g$-modes are indicated, namely the $g_1$, $g_2$, and $g_3$-modes. Evidently, the $g$-mode oscillation frequencies rise sharply with the onset of $\Delta$ resonances as neutron stars reach certain critical masses. The $g$-mode frequencies for neutron stars with $\Delta$ resonances are typically much larger than those without, which is attributed to the drastic increase for differences between the two sound speeds as indicated in Fig.~\ref{Fig:T3}. We also present the two pressure modes in Fig.~\ref{Fig:T5}, i.e., the $f$- and $p_1$-modes. The frequencies of the pressure modes also increase after the onset of $\Delta$ resonances, but the amplitude is smaller compared to that of the $g$-modes.

As the frequencies of pressure modes (kHz) are much higher than those of the $g$-modes ($\sim$0.1 kHz), they can only be excited in the late stages of binary neutron star inspiral. The time window of this stage is extremely short, and higher-order oscillation frequencies will not even be excited prior to the binary merger, which are thus not presented here. In contrast, the excitation timescale for the $g$-modes is much longer during inspiral and possesses unique characteristics. In such cases, the $g$-modes could be excited and observing their frequencies could not only identify the presence of $\Delta$ resonances inside neutron star, but also unveil the in-medium $\Delta$ masses and coupling constants. This makes the $g$-mode oscillations a promising tool to probe the properties of neutron star matter.

Note that for a neutron star hosting a first-order phase transition, special treatment is required at the interface of the two phases when estimating the oscillation frequencies, e.g., those predicted by adopting $m_\Delta = 1112 \text{ MeV}$ and $g_{\rho\Delta} = 0$. In particular, we assume that the phase transition at the interface between nuclear matter and $\Delta$ admixed matter is slow, i.e., the oscillation timescale is much shorter than the phase transition timescale. For such scenarios, the stellar interior oscillation Eqs.~(\ref{eq:37}) and (\ref{eq:38}) must be divided into two parts: from the center to the interface, and from the interface to the stellar surface. The matching conditions at the interface between nuclear matter and $\Delta$ admixed matter is~\cite{Finn1987_MNRAS227-265-293, Sotani2011_PRD83-024014, Miao2024_TAJ964-31}:
\begin{eqnarray}
W_{+} &=& W_{-}     \label{eq:W}
\\
V_{+} &=& \frac{\varepsilon_{-}+p_\mathrm{t}}{\varepsilon_{+}+p_\mathrm{t}}V_{-} + \frac{\varepsilon_{-}-\varepsilon_{+}}{\varepsilon_{+}+p_\mathrm{t}}\cdot\frac{e^{2 \Phi-\Lambda}}{\omega^{2} r_\mathrm{t}^{2}}\frac{\mbox{d}\Phi}{\mbox{d}r}W_\mathrm{t}    \label{eq:V}
\end{eqnarray}
where, $r_\mathrm{t}$ represents the position of the density discontinuity point, while $W_\mathrm{t}$ and $p_\mathrm{t}$ represent the values of $W$ and $p$ at the interface ($W_\mathrm{t} = W_{+} = W_{-}$). The quantities with a ``$+$" sign indicate the values on the side with lower density (nuclear matter) at the interface, and these with a "$-$" sign indicate the values on the side with higher density ($\Delta$ admixed matter) at the interface.

\begin{figure}
\centering
\includegraphics[width=\linewidth]{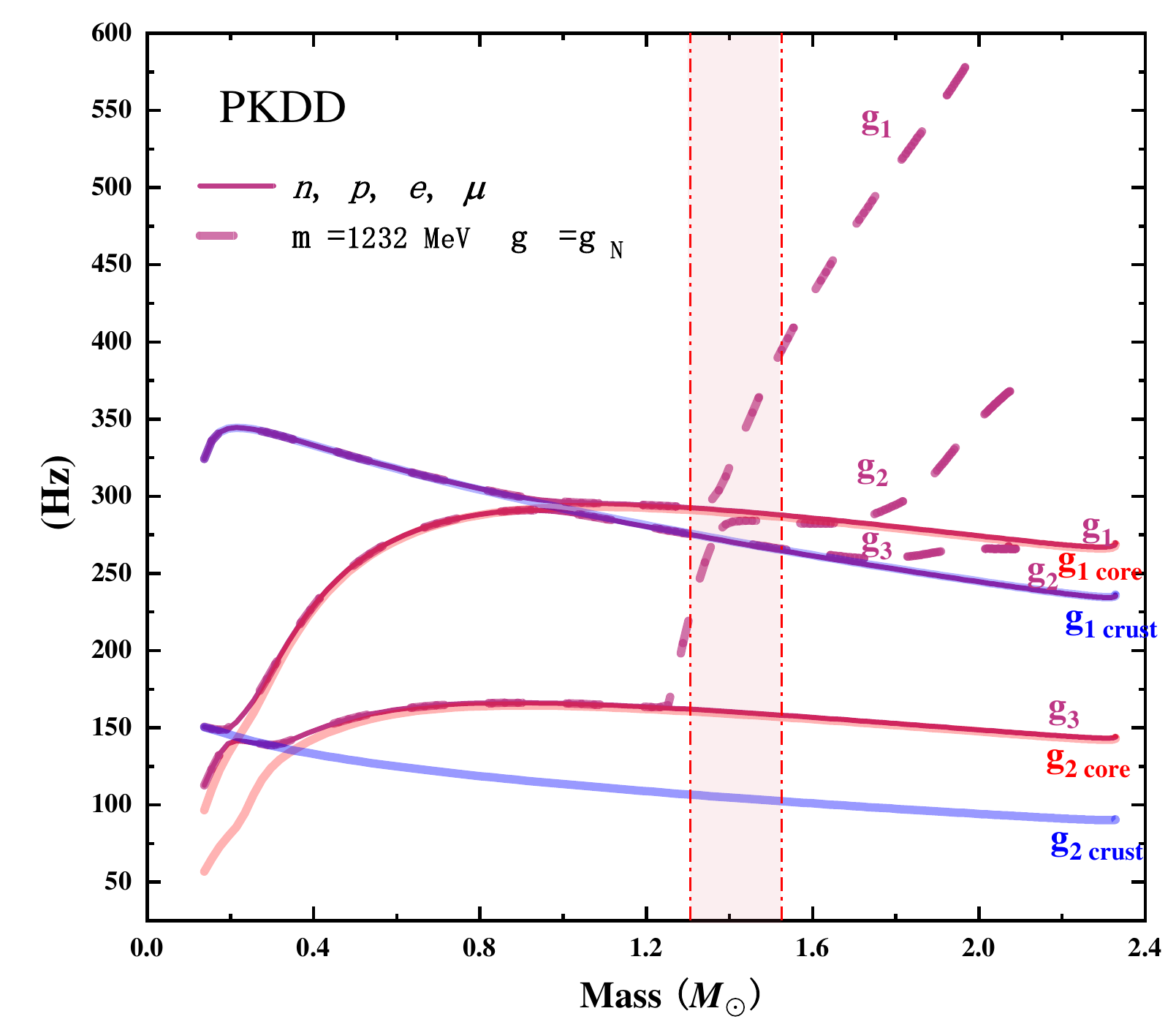}
\caption{\label{Fig:T7} $g$-mode frequencies for traditional neutron stars ($g_1$, $g_2$, $g_3$) and $\Delta$ admixed neutron stars ($g_{1\Delta}$, $g_{2\Delta}$, $g_{3\Delta}$) predicted by adopting $m_\Delta = 1232 \text{ MeV}$ and $g_{\rho\Delta} = g_{\rho N}$, where the nuclear RDF PKDD is employed. The crust $g$-modes ($g_{1\text{crust}}$, $g_{2\text{crust}}$) and core $g$-modes ($g_{1\text{core}}$, $g_{2\text{core}}$) of traditional neutron stars are indicated as well. The vertical red shaded region corresponds to the mass range for $g_{1\Delta}$ examined in Fig.~\ref{Fig:T8}.}
\end{figure}

According to Fig.~\ref{Fig:T5}, there exist several turning points for the frequencies of $g_1$, $g_2$, and $g_3$-modes as functions of neutron star mass. To better understand this phenomenon, we select those predicted by employing the nuclear RDF PKDD as plotted in Fig.~\ref{Fig:T7}, where the first three $g$-mode frequencies for traditional neutron stars ($g_1$, $g_2$, $g_3$) and $\Delta$ admixed neutron stars ($g_{1\Delta}$, $g_{2\Delta}$, $g_{3\Delta}$) obtained with $m_\Delta = 1232 \text{ MeV}$ and $g_{\rho\Delta} = g_{\rho N}$ are presented. The oscillation frequencies for the crusts and cores of neutron stars are also presented, where the first two core $g$-modes $g_{1\text{core}}$ and $g_{2\text{core}}$ are indicated by the light red solid lines and the first two crust $g$-modes $g_{1\text{crust}}$ and $g_{2\text{crust}}$ by blue solid lines. From Fig.~\ref{Fig:T7}, the avoided crossing phenomenon can be identified easily. D. Gondek detailed the reasons for the avoided crossing phenomenon in $f$- and $p$-modes, which arise from the different EOSs and oscillation properties between the stellar core and crust~\cite{Gondek1999_AAP344-117}. For $g$-modes of $\Delta$ admixed neutron stars indicated by the dashed curves, this phenomenon could be attributed to the different properties of matter in the outer core ($npe\mu$ matter) and inner core ($npe\mu\Delta$ matter). For neutron stars with the mass $M < 1.2 M_\odot$, the frequencies of $g_n$ and $g_{n\Delta}$ modes overlap perfectly as the $\Delta$ resonance has not yet emerged. For traditional neutron stars, the squared difference in sound speeds forms two peaks in the core and crust regions, which forms two weakly coupled resonance cavities. In such cases, the crust $g$-mode oscillations ($g_{1\text{crust}}$ and $g_{2\text{crust}}$) dominate in low-mass neutron stars. As the stellar mass increases, the crust $g$-mode frequencies decrease while the core $g$-mode frequencies ($g_{1\text{core}}$ and $g_{2\text{core}}$) increase and take dominance, thereby triggering the avoided crossing phenomenon.

Our previous study~\cite{Sun2025_PRD111-103019} has explored this mechanism in detail, and the present work further extends it to systems containing $\Delta$ admixed matter. As neutron star's mass increases, various $\Delta$ baryons and hyperons emerge in the core region. Their appearance causes a significant increase in $\Delta c^2$, which in turn leads to an elevation in the $g$-mode frequencies. These regions with large $\Delta c^2$, namely the $\Delta$ admixed matter, forms an additional resonance cavity in the inner core. When the mass is sufficiently large, the oscillation energy transfers from the outer core to the inner core, and the overall oscillation frequency is determined by the properties of the inner core. This explains the sudden increase in frequency for neutron stars around $1.4 M_\odot$ as shown in Fig.~\ref{Fig:T7}.

\begin{figure}
\centering
\includegraphics[width=\linewidth]{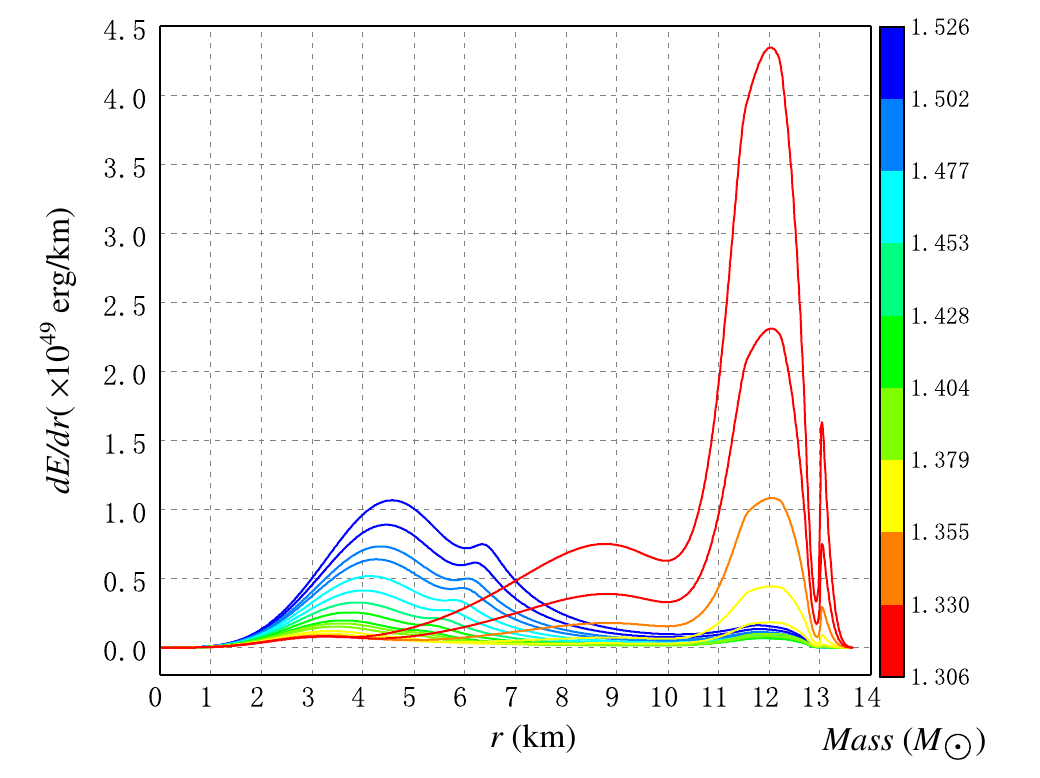}
\caption{\label{Fig:T8}Radial distribution of the oscillation energy $\mbox{d}E/\mbox{d}r$ for the $g_{1\Delta}$ mode oscillation inside neutron stars, corresponding to the vertical red shaded region in Fig.~\ref{Fig:T7}. The transition from red to blue indicates a change in the neutron star's mass from $1.306 M_\odot$ to $1.526 M_\odot$.}
\end{figure}

To investigate the energy transfer from the outer core to the inner core, we examine the radial distribution of the oscillation energy for neutron stars with masses between $1.306 M_\odot$ and $1.526 M_\odot$ (corresponding to the shaded region in Fig.~\ref{Fig:T7}). Under the Cowling approximation, the radial energy distribution for a given eigenmode can be expressed as~\cite{Reisenegger1992_Apj395-240-249, McDermott1983_ApJ268-837, Zheng2023_PRD107-103048}
\begin{equation}
\frac{\mbox{d}E}{\mbox{d}r}=\frac{\omega^{2}}{2}(p+\varepsilon)e^{\Lambda-\Phi}\left[\frac{W^{2}}{r^{2}}+\ell(\ell+1)V^{2}\right].
\label{eq:20}
\end{equation}
Here we primarily focus on the $g_{1\Delta}$ mode, which possesses the largest energy and is likely to be observed. Fig.~\ref{Fig:T8} illustrates the profiles of $\mbox{d}E/\mbox{d}r$ for the first $g$-mode in the neutron stars with $\Delta$ admixed cores obtained by adopting $m_\Delta = 1232 \text{ MeV}$ and $g_{\rho\Delta} = g_{\rho N}$, where the neutron star mass increases as the color changes from red to blue. For a $1.306 M_\odot$ star, the crust-core boundary is located at approximately $r = 12.8 \text{ km}$. A portion of the oscillation energy still resides in the crust of the neutron star, but the primary contribution comes from the outer core. As the mass increases, the oscillation energy gradually shifts inward, and the crustal contribution vanishes. For a $1.526 M_\odot$ star, the oscillation energy is mainly concentrated in the inner core, and the frequency is determined by the $\Delta$ admixed matter in the inner core. If we adopt the nuclear RDF DD-ME2, similar behavior is observed, albeit with subtle differences: their first $g$-mode frequencies transition directly from being crust-dominated to core-dominated. Note that such types of frequency jumps are not unique and may also take place in hyperonic cores~\cite{Tran2023_PRC108-015803} and quark cores~\cite{Constantinou2021_PRD104-123032}, all of which originate from the increase in $\Delta c^2$ associated with phase transitions.

\begin{figure}
\centering
\includegraphics[width=\linewidth]{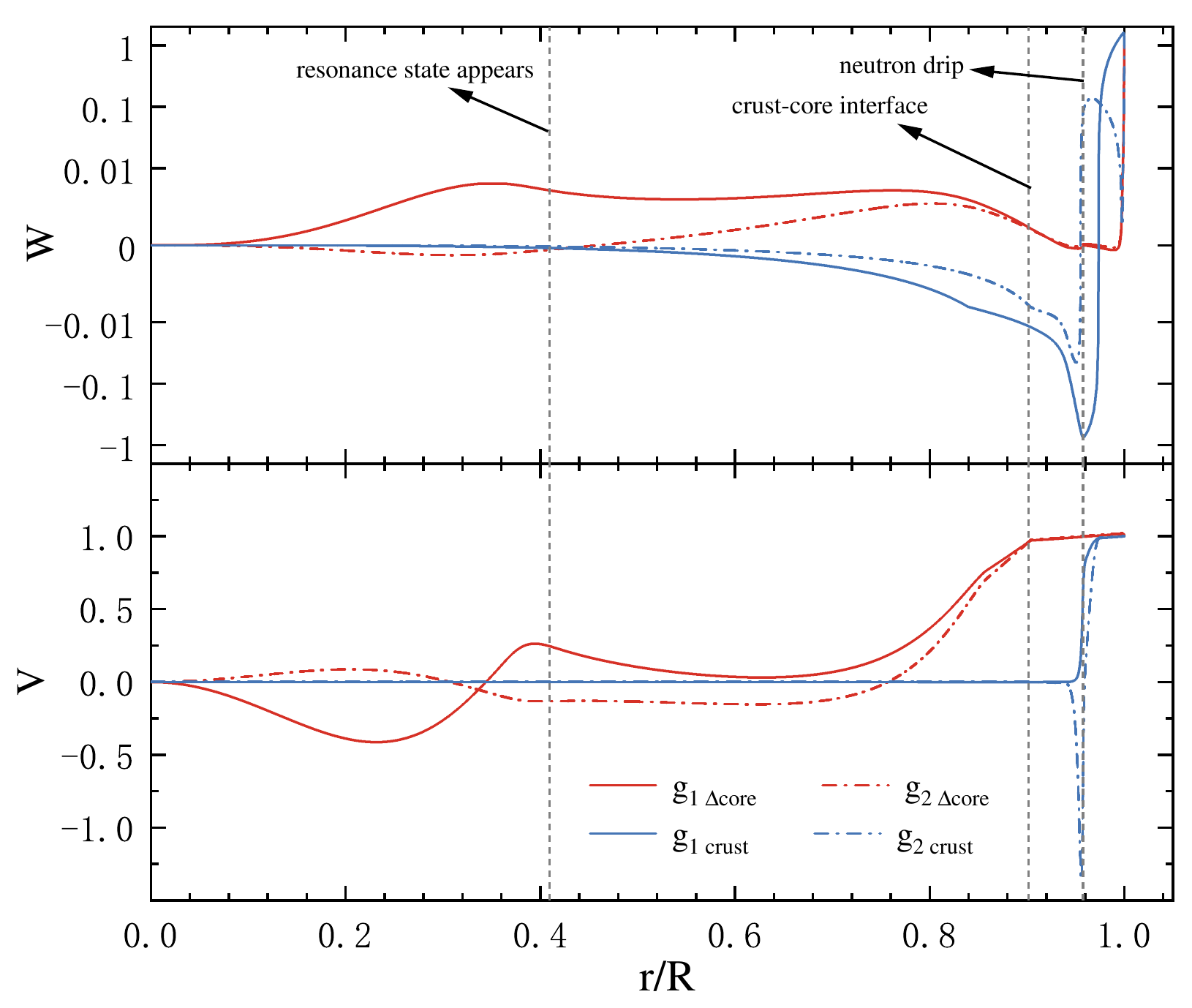}
\caption{\label{Fig:T11}Normalized radial ($W$) and angular ($V$) perturbation amplitudes of $g$ mode oscillations in the $1.4\ M_\odot$ neutron star corresponding to those indicated in Fig.~\ref{Fig:T7} and Fig.~\ref{Fig:T8}. The blue solid and dashed lines represent the $g_\mathrm{1crust}$ and $g_\mathrm{2crust}$ modes in neutron stars' crusts, the red solid and dashed lines represent the $g_{1\Delta\mathrm{core}}$ and $g_{2\Delta\mathrm{core}}$ modes of the core containing $\Delta$ resonances. The positions of neutron drip, crust-core interface, and onset of $\Delta$ resonances are indicated by the vertical dashed lines.}
\end{figure}

Note that for cold neutron stars the crust $g$-mode oscillations could be hindered by the elasticity~\cite{Gao2025_PRD112-123006} and superfluidity~\cite{Gusakov2013_PRD88-101302}.
However, these factors are neglected in our current study, so that the crust $g$-modes persist. In Fig.~\ref{Fig:T11}, we present the normalized radial and angular perturbation amplitudes $W$ and $V$ for the $g$-modes of a $1.4\ M_\odot$ neutron star, where $\Delta$ resonances are included by adopting $m_\Delta = 1232 \text{ MeV}$ and $g_{\rho\Delta} = g_{\rho N}$ as well as the nuclear RDF PKDD. Here $g_{1\Delta\text{core}}$ denotes the core $g$-mode containing $\Delta$ resonances, which is obtained by setting $\Delta c^2=0$ in the crust and differs from $g_{1\text{core}}$ in Fig.~\ref{Fig:T7}. For the amplitudes of the crust $g$-modes, i.e., $g_{1\text{crust}}$ and $g_{2\text{crust}}$, there exists an obvious jump at the position of the neutron drip, and the amplitudes of the radial and tangential eigenfunctions are mainly concentrated in the outer layer of the crust, which is also the main region for outer crust $g$-mode oscillations. For the core $g$-modes containing $\Delta$ resonances, the amplitudes of the eigenfunctions for $g_{1\Delta\text{core}}$ and $g_{2\Delta\text{core}}$ modes are relatively small compared with the crust $g$ modes. Withe the emergence of $\Delta$ resonances in the inner core, the oscillations becomes evident, in which the energy of the core $g$-modes is concentrated.

\begin{figure}
\includegraphics[width=\linewidth]{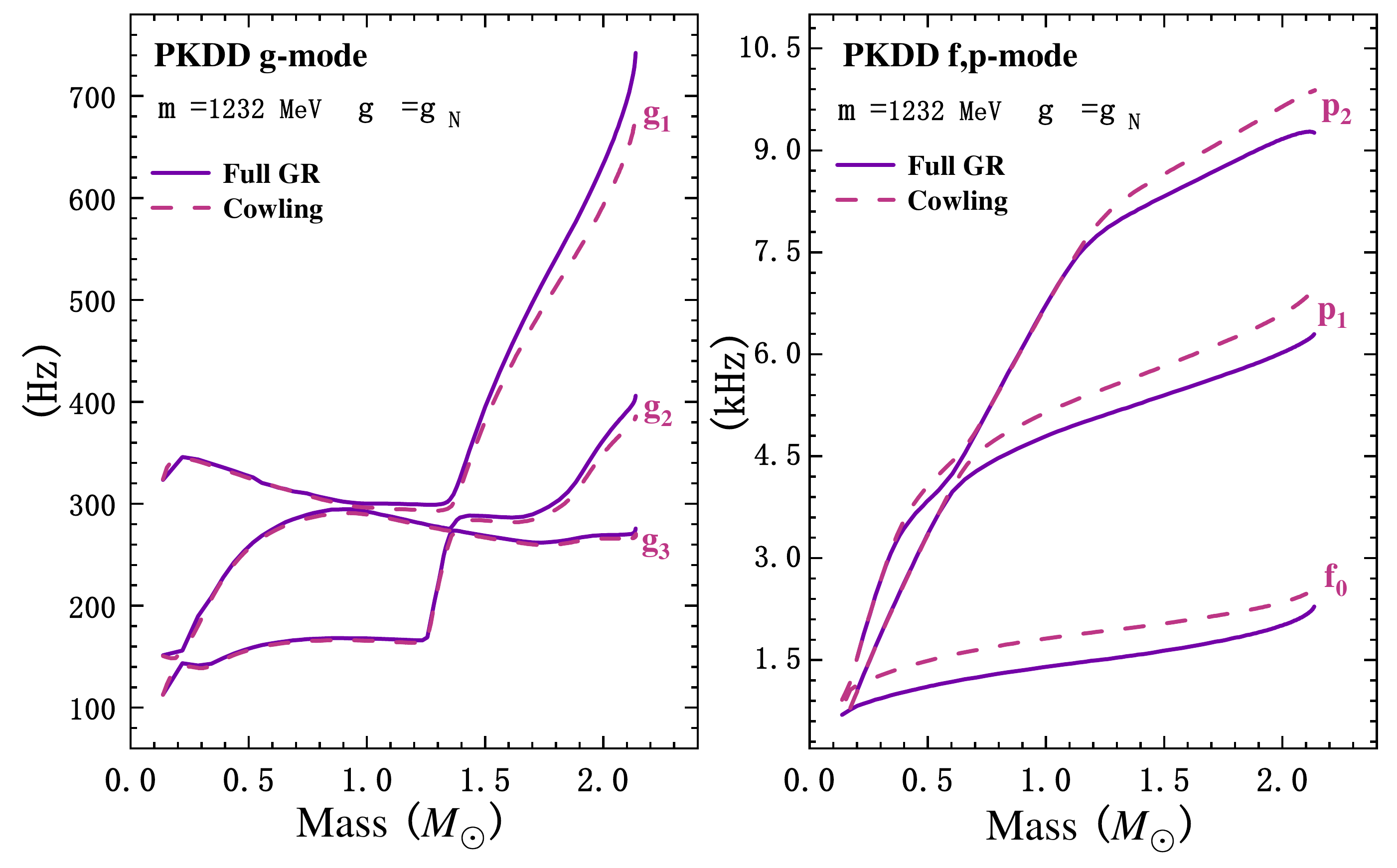}
\caption{\label{Fig:T9}Frequencies of $g$, $f$, and $p$-mode oscillations ($\nu$) as functions of neutron star mass $M$, which are estimated under the  full General Relativity (solid) and Cowling approximation (dashed). The left panel illustrates the first three oscillation frequencies of the $g$-mode, while the right one presents the oscillation frequencies of  $f$, $p_1$, and $p_2$ modes.}
\end{figure}

\begin{table*}[t]
\centering
\caption{\label{Tab:Table 3}Frequencies of $g_1$ and $f$ modes for $1.4\ M_\odot$ and $2\ M_\odot$ neutron stars estimated with two approaches, i.e., employing full general relativity ($\nu_\mathrm{GR}$) with spacetime perturbations and Cowling approximation ($\nu_\mathrm{Cowling}$) without spacetime perturbations, where the percentage error (P.E.) of introducing the Cowling approximation is indicated as well.}
\begin{tabular}{c|c|c|c|cc|c|cc|c}
\hline\hline
   Nuclear RDF    & $m_\Delta$ &  $\alpha_{\rho\Delta}$     & Mass     & $\nu_\mathrm{GR}(g_1)$     & $\nu_\mathrm{Cowling}(g_1)$   & P.E.   & $\nu_\mathrm{GR}(f_1)$     & $\nu_\mathrm{Cowling}(f_1)$   & P.E. \\
           &  MeV       &                            & $M_{\odot}$     & Hz        & Hz         & $\%$       & Hz        & Hz         & $\%$ \\ \hline
\multirow{8}*{PKDD} & \multirow{4}*{1112} & \multirow{2}*{0}
                        & $1.4$    & $668.28$      &  $669.89$       & $0.24$       & $1938.98$      &  $2476.05$       & $27.70$ \\
 & &                    & $2$    & $583.10$      &  $577.25$      & $1.00$          & $2211.03$      &  $2618.29$       & $18.42$ \\ \cline{3-10}
& &   \multirow{2}*{1}  & $1.4$    & $854.75$      &  $862.01$      & $0.85$     & $1913.68$      &  $2378.05$       & $24.27$ \\
 & &                    & $2$    & $814.21$      &  $805.32$      & $1.09$       & $2316.84$      &  $2675.18$       & $15.47$ \\ \cline{2-10}
 & \multirow{4}*{1232} & \multirow{2}*{0}
                        & $1.4$    & $725.86$      &  $715.43$       & $1.44$     & $1785.63$      &  $2215.29$       & $19.40$ \\
 & &                    & $2$    & $691.58$      &  $673.64$      & $2.59$        & $2093.68$      &  $2475.59$       & $18.24$\\ \cline{3-10}
 & & \multirow{2}*{1}   & $1.4$    & $326.93$      &  $323.16$       & $1.15$     & $1576.92$      &  $1995.51$       & $26.54$ \\
 & &                    & $2$    & $635.26$      &  $592.78$      & $6.69$        & $2004.21$      &  $2350.42$       & $17.27$\\ \hline
\multirow{8}*{DD-ME2} & \multirow{4}*{1112} & \multirow{2}*{0}
                        & $1.4$    & $533.85$      &  $531.36$       & $0.47$     & $1834.62$      &  $2371.90$       & $29.29$ \\
 & &                    & $2$    & $464.49$      &  $456.44$      & $1.73$        & $2023.08$      &  $2449.66$       & $21.09$\\ \cline{3-10}
& &   \multirow{2}*{1}  & $1.4$    & $678.01$      &  $647.31$       & $4.53$     & $1876.71$      &  $2372.34$       & $26.41$ \\
 & &                    & $2$    & $606.84$      &  $576.86$      & $4.94$        & $2075.26$      &  $2483.97$       & $19.69$\\ \cline{2-10}
 & \multirow{4}*{1232} & \multirow{2}*{0}
                        & $1.4$    & $487.13$      &  $473.02$       & $2.90$     & $1689.25$      &  $2135.08$       & $26.39$ \\
 & &                    & $2$    & $514.74$      &  $494.48$      & $3.94$        & $1930.53$      &  $2303.37$       & $19.31$\\ \cline{3-10}
 & & \multirow{2}*{1}   & $1.4$    & $460.20$      &  $447.79$       & $2.70$     & $1633.89$      &  $2080.71$       & $27.35$ \\
 & &                    & $2$    & $583.16$      &  $555.07$      & $4.82$        & $1912.63$      &  $2285.62$       & $19.50$\\ \hline
\end{tabular}
\end{table*}

The non-radial oscillations illustrated above are obtained based on the Cowling approximation. Although its validity has been widely verified, to show the variations caused by introducing Cowling approximation, we perform full general relativity calculations for the $g$, $f$, and $p$-modes of neutron stars with $\Delta$ resonance ($m_\Delta = 1232 \text{ MeV}$, $g_{\rho\Delta} = g_{\rho N}$). Our results are then presented in Fig.~\ref{Fig:T9}, which are compared with those indicated in Fig.~\ref{Fig:T5} adopting Cowling approximation. From the left panel of Fig.~\ref{Fig:T9}, it is evident that the $g_1$, $g_2$, and $g_3$-mode frequencies obtained by the two methods coincide with each other, but with slight deviations once the neutron star mass approaches $2\ M_\odot$, which confirms the reliability of the Cowling approximation for $g$-mode frequencies. In the right panel, the frequencies of the $f$, $p_1$, and $p_2$-modes are indicated. The pressure modes exhibit an avoided crossing phenomenon between crust-dominated and core-dominated oscillations~\cite{Gondek1999_AAP344-117}. The frequencies of the pressure modes ($f$, $p_1$, and $p_2$-modes) obtained by the two methods are almost identical for the crust modes, but with large differences for the core modes. This is because the crust contributes very little to the total mass of the star and exhibits weak relativistic effects, while the core has a larger mass with stronger relativistic effects, thus leading to increased deviations when Cowling approximation is employed. Table~\ref{Tab:Table 3} lists the mode frequencies and percentage errors of the Cowling approximation with respect to the full general relativity scenarios for $1.4\ M_\odot$ and $2\ M_\odot$ neutron stars, which are consistent with the findings in the references~\cite{Zhao2022_PRD105-103025, Yoshida1997_MNRAS117-122-289, Sotani2020_PRD102-063025}, i.e., the errors for $g$-modes are within 10\%, $f$-modes are approximately 20\%. The high consistency of the $g$-modes indicates that the relativistic effects for these oscillations are relatively weak, hence their gravitational wave signatures are more difficult to detect, demanding more sensitive detectors for future observations.


\section{\label{sec:res}Conclusion and prospect}

In this work, we investigate systematically the impact of hyperons and $\Delta$ resonances on the non-radial oscillations in neutron stars, where the relativistic energy density functionals TW99~\cite{Typel1999_NPA656-331}, PKDD~\cite{Long2004_PRC69-034319} and DD-ME2~\cite{Lalazissis2005_PRC71-024312} are employed for nuclear matter. The
$\omega$-hyperon couplings are taken as $g_{\omega \Lambda} = g_{\omega \Xi} = g_{\omega \Sigma} = g_{\omega N}$ so that the $\phi$-baryon coupling vanishes according to SU(3) symmetry-inspired prescriptions for meson-baryon couplings~\cite{Schaffner1996_PRC53-1416, Weissenborn2012_PRC85-065802, Miyatsu2013_PRC88-015802, Oertel2015_JPG42-075202}, while the
$\sigma$-hyperon couplings are fixed by reproducing the hyperon potential depths in nuclear matter. The universal coupling scheme is employed for the $\Delta$-meson couplings, while we have also adopted various $g_{\rho \Delta}$ and $m_\Delta$ to account for the uncertainty of isovector interactions and the Breit-Wigner mass distribution of the $\Delta$ isobar. It is found that the inclusion of $\Delta$ resonances is essential for neutron stars to accommodate the recent mass and radius measurements of PSR J0030+0451, PSR J0740+6620, PSR J0437-4715, PSR J0614-3329, and HESS J1731-347.

The non-radial oscillations of the corresponding neutron stars are then examined employing two approaches, i.e., full general relativity ($\nu_\mathrm{GR}$) with spacetime perturbations and Cowling approximation ($\nu_\mathrm{Cowling}$) without spacetime perturbations. Our main findings are summarized as follows:
\begin{itemize}
\item The emergence of $\Delta$ resonances in the core significantly increases the $g$-mode frequencies. This enhancement originates from the modification of the internal composition and density gradient caused by the $\Delta$ population, which strengthens the buoyancy frequency. The coexistence of $\Delta$ and hyperons can form a resonant cavity inside the neutron star, allowing the oscillation energy to be redistributed between the inner ($\Delta$ admixed) and outer cores, particularly for massive stars.
\item The $f$-mode frequency also increases in the presence of $\Delta$ resonances, but is less pronounced compared with that of the $g$-mode. This indicates that the $g$-mode is more sensitive to microphysical variations in the stellar interior and can serve as a promising probe for detecting the appearance of exotic components or phase transitions.
\item A comparison between the full general relativistic (GR) treatment and the Cowling approximation shows that the difference in $g$-mode frequencies is less than about 10\%, suggesting the Cowling approximation remains a reliable approach for low-frequency gravity modes. However, the discrepancies for $f$ and $p$ modes can exceed 20\%, implying that a full GR framework is necessary for high-frequency oscillations.
\end{itemize}

The current study provides a new theoretical perspective on the role of $\Delta$ resonances in the dynamics of dense matter and offers potential observational signatures for future neutron star asteroseismology and gravitational-wave studies. In particular, the high sensitivity of the $g$-mode frequencies to the composition of neutron star matter could potentially help to constrain the EOSs and reveal the possible presence of exotic particles inside neutron stars.

\section*{ACKNOWLEDGMENTS}
The authors would like to thank Prof. Lap-Ming Lin for fruitful discussions.  This work was supported by the National Natural Science Foundation of China (Grant No. 12275234).


%

\end{document}